\documentclass[useAMS]{mn2e}
\usepackage{graphicx}
\usepackage{multirow}

\newcommand{\fbqs} {FBQS J1408+3054}
\newcommand{\kms} {km~s$^{-1}$}
\newcommand{\gtrsim} {\ {\raise-.5ex\hbox{$\buildrel>\over\sim$}}\ }

\newcommand{\aliii}{Al\,{\sc iii}}

\newcommand{\civ}{C\,{\sc iv}}

\newcommand{\feii}{Fe\,{\sc ii}}
\newcommand{\feiii}{Fe\,{\sc iii}}

\newcommand{\mgii}{Mg\,{\sc ii}}

\newcommand{\MgII}{Mg\,{\sc ii}\,$\lambda\lambda$2796,2803}

\newcommand{\Nv}{N\,{\sc v}}

\begin{document}

\title[Variable BAL Quasar FBQS J1408+3054]{Implications of Dramatic Broad Absorption Line Variability in the Quasar FBQS J1408+3054}
\author[Hall et al.]{P. B. Hall,$^1$
K. Anosov,$^1$ 
R. L. White$^2$, 
W. N. Brandt$^3$, 
M. D. Gregg$^{4,5}$,
\newauthor
R. R. Gibson$^6$,
R. H. Becker,$^{4,5}$
D. P. Schneider$^3$\\
$^{1}$Department of Physics and Astronomy, York University, Toronto, ON, Canada M3J 1P3\\
$^{2}$Space Telescope Science Institute, Baltimore, MD, USA 21218\\
$^{3}$Department of Astronomy \& Astrophysics, Pennsylvania State University, University Park, PA, USA 16802\\
$^{4}$Department of Physics, University of California Davis, Davis, CA, USA 95616\\
$^{5}$IGPP, Lawrence Livermore National Laboratory, Livermore, CA, USA 94550\\
$^{6}$Department of Astronomy, University of Washington, Seattle, WA, USA 98195 
}

\date{}


\maketitle

\label{firstpage}

\begin{abstract}
We have observed a dramatic change in the spectrum of the formerly heavily
absorbed `overlapping-trough' iron low-ionization broad absorption line
(FeLoBAL) quasar FBQS J1408+3054.
Over a time span of between 0.6 to 5 rest-frame years, the 
\mgii\ trough outflowing at 12,000~km~s$^{-1}$
decreased in equivalent width by a factor of two
and the \feii\ troughs at the same velocity disappeared.
The most likely explanation for the variability is that a structure in the
BAL outflow moved out of our line of sight to the ultraviolet continuum
emitting region of the quasar's accretion disk.  
Given the size of that region, this structure must have 
a transverse velocity of between 2600~km~s$^{-1}$ and 22,000~km~s$^{-1}$.
In the context of a simple outflow model, we show that this BAL structure 
is located between approximately 
5800 and 46,000 Schwarzschild radii from the black hole.
That distance corresponds to 1.7 to 14~pc, 11 to 88 times farther from 
the black hole than the H$\beta$ broad-line region.
The high velocities and the parsec-scale distance for at least this one 
FeLoBAL outflow mean that
not all FeLoBAL outflows can be associated with galaxy-scale outflows in
ultraluminous infrared galaxies transitioning to unobscured quasars.
The change of FBQS J1408+3054 from an FeLoBAL to a LoBAL quasar also means that 
if (some) FeLoBAL quasars have multiwavelength properties which distinguish 
them from HiBAL quasars, then some LoBAL quasars will share those properties.
Finally, we extend previous work on how multiple-epoch spectroscopy of BAL and
non-BAL quasars can be used to constrain the average lifetime of BAL episodes
(currently $>$60 rest-frame years at 90\% confidence).
\end{abstract}



\section{Introduction}	\label{intro}

Broad Absorption Line (BAL) quasars are a subset of quasars which exhibit
strong absorption troughs from gas flowing outwards with velocities of
thousands to tens of thousands of km~s$^{-1}$ (e.g., Allen et al. 2010).
The traditional definition is that BAL quasars are those 
which have \civ\ absorption which dips at least 10\% below the continuum level
for a contiguous stretch of at least 2000~\kms\ at
blueshifts of $\geq$3000~\kms\ relative to the quasar (Weymann et al. 1991).
Other proposed definitions (Hall et al. 2002; Trump et al. 2006; Gibson et al. 
2009) relax some of the conditions of the traditional definition to include
more cases of potential intrinsic absorption with the tradeoff of increased
contamination by blended narrow absorbers (Knigge et al. 2008) whose origin(s),
especially at low outflow velocities, may differ from those of BAL outflows.

BAL quasars are often classified into three subtypes:
high-ionization (HiBAL), low-ionization (LoBAL) and iron LoBAL (FeLoBAL).  
HiBAL quasars show absorption from only
relatively high-ionization species such as \civ\ and \Nv.  
LoBAL quasars show high-ionization features plus
absorption from low-ionization species such as \mgii\ and \aliii. 
`Iron LoBAL' (Becker et al. 1997) or `FeLoBAL' (Becker et al. 2000) 
quasars show high- and low-ionization troughs 
plus absorption from excited states of \feii, \feiii, or both.
There is no commonly accepted definition for how broad or deep the \mgii\ or
\feii\ absorption has to be for a quasar to be classified as a LoBAL or FeLoBAL
quasar (Trump et al. 2006; Gibson et al. 2009; Zhang et al. 2010).
However, such troughs should at least be broader than intervening 
\mgii\ absorption systems, the vast majority of which have widths 
$<$500~\kms\ (Mshar et al. 2007).  
A minimum depth of 10\% is appropriate for LoBAL and FeLoBAL troughs just as 
for HiBAL troughs, though in all cases this is not a physical limit but
reflects the limitations imposed by spectral signal-to-noise ratios
and our inability to model quasar continua exactly.

BAL troughs are often highly saturated, so that the trough depth is determined
less by the gas column density and more by the covering fraction; that is,
the fraction of the continuum source occulted by the absorbing structure
(e.g., Arav et al. 1999).
Variability in BAL trough strengths is relatively common 
(see the exhaustive reference list in Hall et al. 2007,
and Gibson et al. 2008) but large ($>$50\%) changes in absorption
equivalent width are rare (Crenshaw et al. 2000; Lundgren et al. 2007;
Hamann et al. 2009; Leighly et al. 2009; Gibson et al. 2010).

The $z$=0.848 quasar FBQS J140806.2+305448 (hereafter FBQS J1408+3054) was
first reported by Gregg et al. (1996), where it was misclassified as a star,
and was first identified as a LoBAL quasar in White et al. (2000).  Hall et al.
(2002) described it as a member of the class of `overlapping-trough' FeLoBAL 
quasars because its \feii\ absorption troughs are broad
enough to overlap and yield almost no continuum windows below 2800\,\AA.
However, the covering fraction of its \feii\ absorption was only
$\sim$75\% at its discovery epoch, 
unlike some overlapping-trough objects
which have $\gtrsim$95\% covering fractions.
\fbqs\ also likely has an FR~II radio morphology (Gregg et al. 2006).

In this work we present spectroscopy of FBQS J1408+3054 spanning nearly 8 years
in its rest frame.  The spectroscopic dataset shows mild variability followed
by a dramatic decline in \feii\ and \mgii\ absorption strength
(\S~\ref{data}).  
Substantial trough variability has previously been observed in the
overlapping-trough FeLoBAL quasar SDSS J043742.81$-$004517.6
(Hall et al. 2002), but \fbqs\ is the first known case of an 
FeLoBAL spectrum changing to a LoBAL spectrum.\footnote{We note that the
absorption system D+E in NGC 4151, while too narrow (FWHM = 435 \kms)
and low-velocity (491 \kms) to qualify as a BAL trough, exhibited only
\mgii\ absorption in 1994-1996 but also exhibited strong \feii\ absorption 
in 1999 (Kraemer et al. 2001) and weak \feii\ absorption in 2002 
(Kraemer et al. 2006), indicative of transverse motion 
of an intrinsic quasar absorption system such as that reported herein.}
We estimate a black hole mass for \fbqs\ in \S~\ref{mbh}.
In \S~\ref{discuss} we constrain the BAL structure's size and transverse 
velocity and present a simple outflow model which we use to constrain the 
BAL structure's location and age.  
We also discuss the implications of our observations for the average
episodic lifetime of BAL troughs in quasar spectra and for the link(s) 
between BAL subtypes and between BAL troughs and FR~II radio morphologies.
We end with our conclusions in \S~\ref{con}.

\section{Data}	\label{data}

\subsection{Optical Spectroscopy}

\begin{figure*}
\vspace{-0.4cm}
\makebox[\textwidth]{
\includegraphics[width=0.825\textwidth]{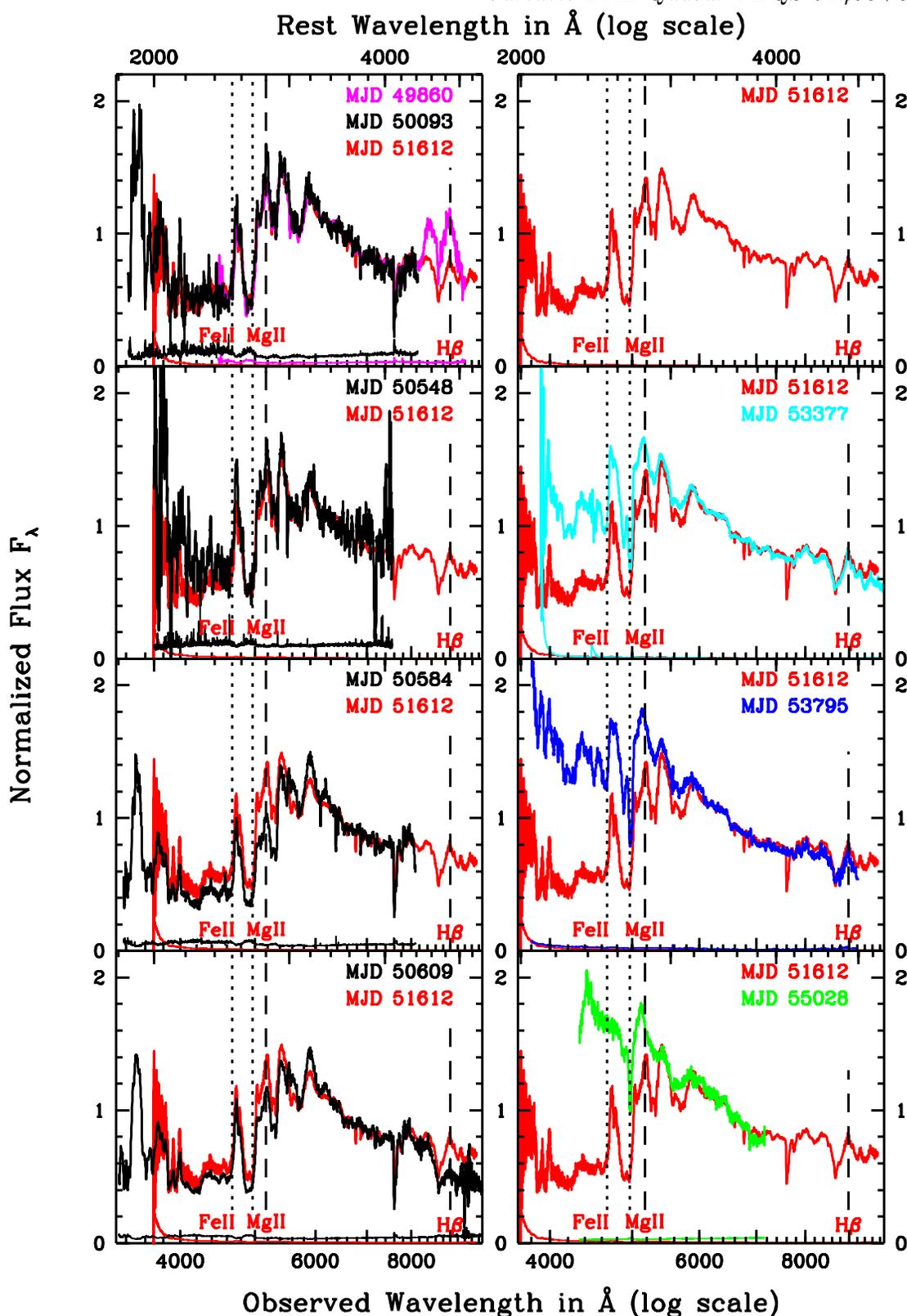}}
\vspace{-0.6cm}
\caption{Spectra and error arrays of FBQS J1408+3054 from 1995 (top left) to 
2009 (bottom right), each smoothed with a seven-pixel boxcar, colour-coded,
and labelled with the MJD of observation.  
For comparison, 
in all panels the MJD 51612 Keck ESI spectrum is plotted in red.
The vertical axes are normalized $F_\lambda$; all spectra have been normalized
to an average of unity in the observed wavelength range 6000-7000~\AA.
The lower horizontal axes are observed wavelengths and
the upper are rest-frame wavelengths at $z_{em}$=0.848, both in \AA.
Dashed vertical lines show the wavelengths of emission from H$\beta$ and \mgii.
Dotted vertical lines show the wavelengths of absorption from
\feii\ $\lambda$2632 (the longest-wavelength component of the UV1 multiplet)
and \MgII\ at $z_{abs}$=0.775, the central redshift of the outflow.
In the left-hand panels, only minor variations in the absorption are seen to
occur.
In the right-hand panels, a strong weakening in the absorption is evident
between the Keck ESI spectra taken on MJD 51612 (red) and on 53377 (cyan).
This weakening continues in the MJD 53795 SDSS spectrum (blue)
and in the MJD 55028 HET LRS spectrum (green).}
\label{fig:RickWhiteFig.eps}
\end{figure*}

\begin{figure*}
\makebox[\textwidth]{
\includegraphics[width=1.15\textwidth]{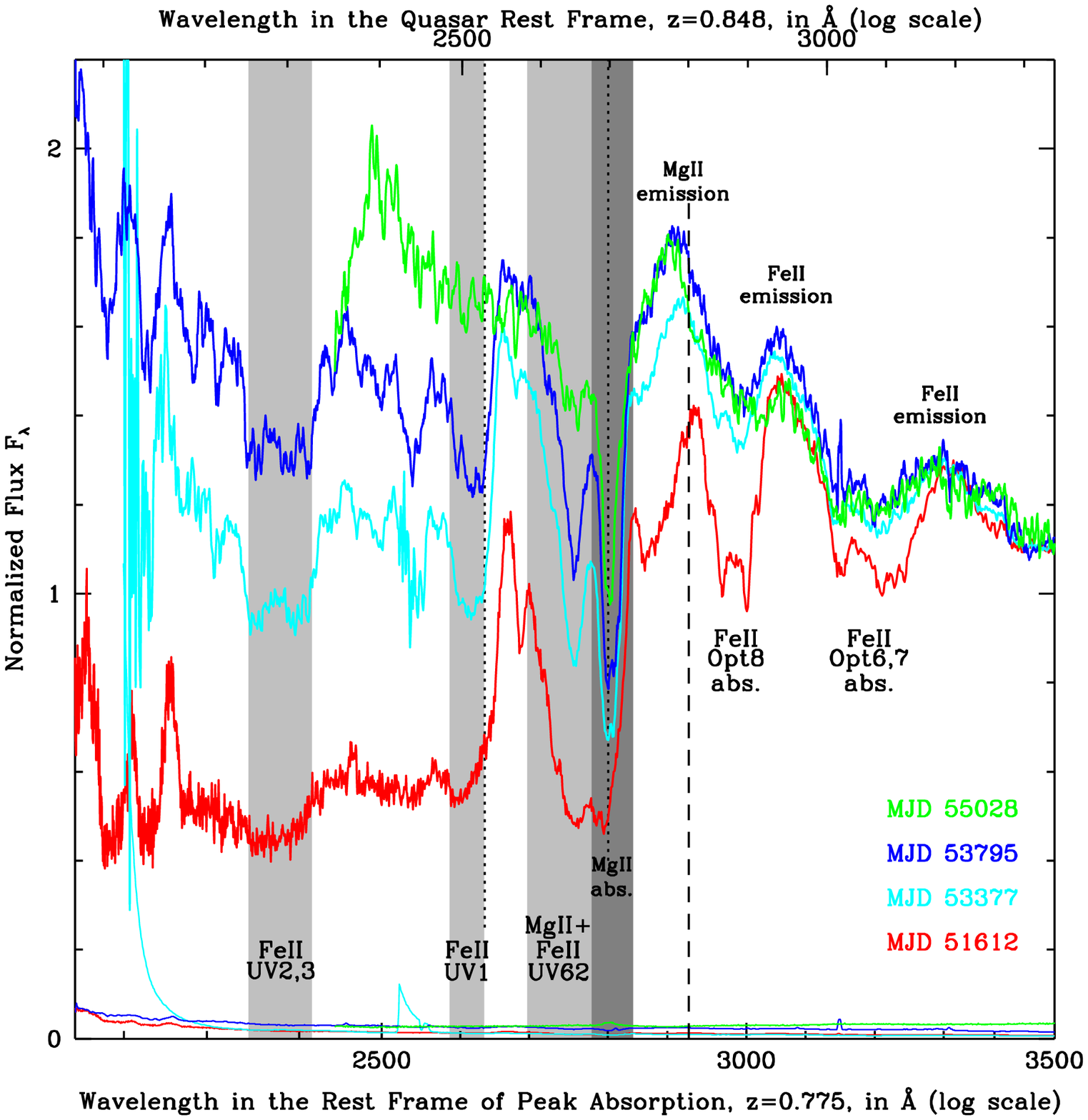}}
\caption{Rest-frame near-ultraviolet spectra and error arrays of
FBQS J1408+3054 during the disappearance of its \feii\ troughs.
The spectra have been smoothed with a seven-pixel boxcar, colour-coded as in
Figure~\ref{fig:RickWhiteFig.eps}, and labelled with the MJD of observation.
The lower horizontal axis 
shows wavelengths in the rest frame of peak absorption ($z$=0.775) and the
upper shows rest-frame wavelengths at the quasar $z_{em}$=0.848, both in \AA.
The vertical dashed and dotted lines are the same 
as in Figure~\ref{fig:RickWhiteFig.eps}.
The dark grey shaded region shows the wavelength range of 
low-velocity \mgii\ absorption (squares in Figure~\ref{fig:EWFig.eps}).
The light grey shaded regions show the wavelength ranges absorbed by
\feii\ multiplets UV1 and UV2+UV3 at $z$=0.775 
and by a blend of \feii\ UV62 and high-velocity \mgii\ absorption.
Regions of absorption by \feii\ multiplets Opt6+Opt7 and Opt8 
are also labelled, as are broad emission lines of \mgii\ and \feii.
Note that the MJD 55028 spectrum shows some weakening of the \feii\ Opt8
emission at $\sim$2950~\AA\ in the quasar rest frame.
}
\label{fig:zoom1408.ps}
\end{figure*}

Spectroscopy was obtained using numerous telescope and instrument
combinations (Table~1), including the Echellette Spectrograph
and Imager (ESI; Sheinis et al. 2002) at the Keck telescope, 
the Kast Spectrograph on the Lick Observatory 3m Shane telescope 
(Miller \& Stone 1993), and the Low Resolution Spectrograph 
(LRS; Hill et al. 1998) on the Hobby-Eberly Telescope (HET). 
Data were obtained, reduced and calibrated using standard procedures
(Gregg et al. 1996; White et al. 2000; Gregg et al. 2002),
except as we now discuss with regard to the HET spectrum.

We obtained a spectrum of \fbqs\ with the LRS on the HET 
with the g2 grism (useful observed wavelength coverage
4335-7155 \AA) and a 1$\arcsec$ slit.  We obtained two spectrograms with a
total exposure of 1510 seconds in 2$\arcsec$ seeing at an average airmass of 
1.26.  A flux calibration star was not observed.  We used the known throughput
of the instrumental configuration to remove the instrumental response
before correcting for interstellar and atmospheric extinction.
When compared to the other normalized spectra shown in Figure~1, the normalized
HET spectrum exhibited a large-scale curvature resulting in a downturn in flux
over several hundred observed \AA\ at both ends of the spectrum.  We believe
this curvature resulted from slit losses due to differential refraction.  
For the slit to include the quasar and a star bright enough to guide on,
the long axis of the slit had to be oriented about 82$^\circ$ from the
parallactic angle, resulting in differential refraction of $\sim$0.75$\arcsec$
over our wavelength range (Table~I of Filippenko 1982, hereafter F82).
Inspection of the results of the two individual exposures obtained showed that
one exposure yielded a more prominent downturn in its spectrum and also
showed more flux from a third object slightly offset from the centre of 
the slit, suggesting that the quasar was not well centred in the slit in
that exposure.  We therefore dropped that exposure from consideration.
To correct for slit losses in the remaining exposure,
we assumed that the target was centred in the slit at a wavelength of 
6500~\AA.  We computed the offset of the target along the narrow axis of the
slit from 4000-7500~\AA\ using Table~I of F82.  We then used equation 8 of F82
to calculate the fraction of the flux captured by the slit as a function of
wavelength and corrected the spectrum for those wavelength-dependent losses.

The panels of Figure~\ref{fig:RickWhiteFig.eps} 
show all of our spectra of \fbqs\ taken between 1995 and 2009, normalized to
an average of unity between observed 6000\,\AA\ and 7000\,\AA.
The dashed vertical lines show the wavelengths of H$\beta$ and \mgii\ emission
in the quasar rest frame.  The dotted vertical lines show absorption from 
\feii\ $\lambda$2632 and \MgII\ at the redshift of maximum absorption
($z$=0.775, a blueshift of 12,000 km~s$^{-1}$ from the quasar rest frame).
Blueward of \feii\ $\lambda$2632, the spectra suffer
overlapping absorption from hundreds of transitions of \feii.
A strong decrease in absorption strength occurred over the 955 rest-frame
days between the two Keck epochs (MJDs 51612, in red, and 53377, in cyan).
The absorption was still dramatically weaker in the Sloan Digital Sky Survey
(SDSS; York et al. 2000) spectrum (in blue) taken on MJD 53795, 123 rest-frame 
days after the second Keck epoch, and made available in the SDSS Seventh Data 
Release (DR7; Abazajian et al. 2009).
The corrected HET spectrum, taken 668 rest-frame days after the SDSS epoch and
shown in green in the lower right panel of Figure~\ref{fig:RickWhiteFig.eps},
reveals \mgii\ absorption even weaker than in the SDSS epoch
and a complete lack of \feii\ UV1 absorption.

Figure~\ref{fig:zoom1408.ps} displays our near-ultraviolet spectra of 
\fbqs\ from the period of weakening absorption.
Wavelengths (in \AA) are given in the rest frame of peak absorption ($z$=0.775)
along the bottom and in the quasar rest frame ($z_{em}$=0.848) along the top.
As in Figure~\ref{fig:RickWhiteFig.eps},
the vertical dashed line shows the wavelength of \mgii\ in the quasar rest
frame defined by the H$\beta$ line (in the more recent spectra,
the \mgii\ peak is blueshifted relative to the H$\beta$ rest frame)
and the rightmost vertical dotted line 
shows the wavelength of \mgii\ at the velocity of peak absorption, while
the leftmost dotted line shows absorption at the same velocity from \feii\ 
$\lambda$2632, the longest-wavelength component of the \feii\ UV1 multiplet
(Moore 1950). 
The left two light grey shaded regions show the wavelength ranges absorbed by
\feii\ multiplets UV2+UV3 and UV1 at $z$=0.775; as expected,
\feii\ $\lambda$2632 traces the long-wavelength edge of the 
UV1 absorption trough.  
Absorption outside the shaded regions arises predominantly from
hundreds of excited \feii\ transitions whose troughs overlap in velocity space.

The dark grey shaded region shows the wavelength range 
spanned by \mgii\ absorption at velocities $v$=8700~\kms\ to $v$=14,200~\kms.
The rightmost light grey shaded region shows the wavelength ranges absorbed
by a blend of \feii\ UV62 + higher-velocity \mgii\ absorption.  This
high-velocity wing of the \mgii\ trough extends to 2623~\AA\ in the quasar rest
frame ($v$=19,500~\kms), as seen in the HET spectrum from MJD 55028 (green).
The absorption at those wavelengths in the HET spectrum cannot be \feii\ UV62 
because the \feii\ UV62 multiplet is intrinsically weaker than the
UV1 multiplet and the latter is not present in the HET spectrum.  
Without detailed modeling,
it is impossible to be sure that the \feii\ absorption spanned the same 
velocity range as \mgii, but the shape of the MJD 51612 spectrum is
consistent with \feii\ UV62 absorption also extending to 19,500~\kms.

\begin{figure}
\includegraphics[width=84mm]{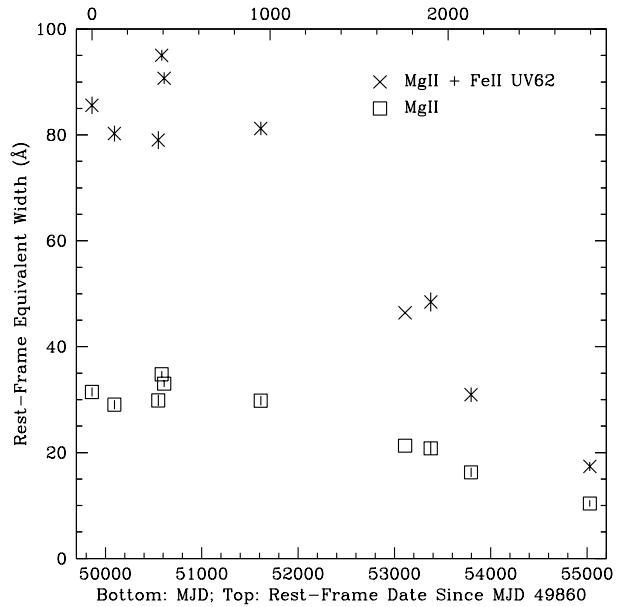}
\caption{Rest-frame equivalent widths in \AA\ of 
\mgii+\feii\,UV62 (crosses)
and low-velocity \mgii\ (squares).
Error bars shown for each point do not include the uncertainty
in continuum placement.
Around rest-frame day 400,
the absorption in both species increased in strength quickly 
(over 19.5 rest-frame days) and then declined gradually (over 13.5 days).
Between rest-frame days 948 and 1903, the \mgii+\feii\,UV62 $W_r$ 
fell by nearly half.  (The points without error bars at day 1759 
are photometric estimates; see \S~\ref{Multi}.)
By rest-frame day 2797, the \mgii+\feii\,UV62 blend contains
only low- and high-velocity \mgii; \feii\,UV62 has vanished.}
\label{fig:EWFig.eps}
\end{figure}

Figure~\ref{fig:EWFig.eps} shows the \mgii\ and blended 
\mgii+\feii\,UV62 rest-frame equivalent widths $W_r$ for all spectra
in Figure~\ref{fig:RickWhiteFig.eps}.
The equivalent widths were measured relative to a continuum established using 
the normalized SDSS spectrum shown in Figure~1.
In that spectrum, a linear continuum in $F_\lambda$ vs. $\lambda$ was 
drawn between the average fluxes at rest-frame 
2550-2600~\AA\ and 3765-3820~\AA.  
In all spectra,
\mgii\ $W_r$ values were measured between 2670-2720~\AA\ and 
\mgii+\feii\,UV62 $W_r$ values were measured between 2585-2720~\AA. 
While there is some uncertainty in the actual continuum level and thus in the
measured $W_r$, the changes in $W_r$ in Figure~\ref{fig:EWFig.eps} 
are much larger than those uncertainties.
As the \mgii+\feii\,UV62 blend consists only of high-velocity \mgii\ in the HET
epoch, we conservatively estimate that the $W_r$ of \mgii\ had decreased
by a factor of two by the HET epoch.

\subsection{Multiwavelength Photometry}	\label{Multi}

In Table~2, we present time-ordered
observed and synthesized photometric magnitudes and colours of \fbqs.
%
The photographic magnitudes listed are the
average of the USNO-B1.0 (Monet et al. 2003) and GSC2.3.2 (Lasker et al. 2008)
magnitudes whenever available and are typically uncertain by $\pm$0.4 
(as compared to $\pm$0.02 for SDSS magnitudes).
The conversions to SDSS 
magnitudes were made using
equation 2 of Monet et al. (2003) except for the Palomar Quick $V$
which used Table 7 of Smith et al. (2002).
In five spectroscopic epochs, the wavelength coverage
was sufficient to enable synthesis of $g-r$ and $r-i$ colours using the SDSS
system response curves of Doi et al. (2010)
but slit losses prevented the synthesis of accurate magnitudes.
Only in the SDSS spectroscopic epoch could both colours and
magnitudes be reliably synthesized.
Corrections for Galactic extinction have not been applied to these magnitudes.

Despite the uncertainties in the 
magnitudes in Table~2,
the $g-r$ colours obtained from them provide good evidence that 
the $g$-band absorption in \fbqs\ was approximately constant in strength 
for at least 27 rest-frame years prior to its weakening sometime 
after MJD 51612. 
However, we cannot rule out the disappearance and reappearance
of an \feii\ BAL structure during that time, especially because
there is a gap in the historical photometric record longer than twice the
timescale over which the \feii\ absorption was later observed to vanish. 

In the two SDSS photometric epochs, 
there was an 18\% increase in $g$-band flux relative to 
the $i$-band flux between MJD 53110 and 53795 (371 rest-frame days).
Assuming that flux increase was uniform across the $g$ band,
we can predict that at MJD 53110 (rest-frame day 1759)
the \mgii+\feii\ trough had $W_r=46$~\AA\ and the \mgii\ trough $W_r=21$~\AA.
These points are plotted in Figure~\ref{fig:EWFig.eps} without error bars.
They suggest that the change in $W_r$ values during the weakening of the BAL
troughs may have been more gradual than rapid, and perhaps not monotonic.

\begin{figure}
\includegraphics[width=84mm]{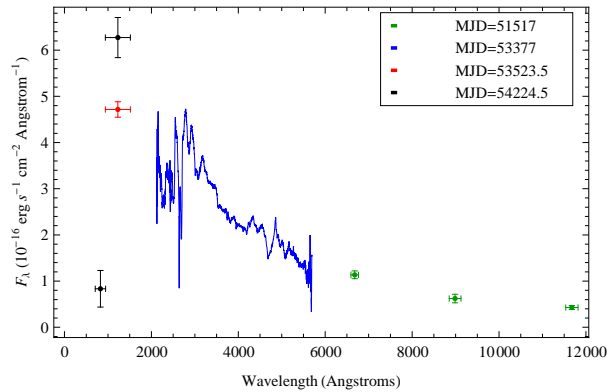}
\caption{Rest-frame ultraviolet/optical spectrum and rest-frame
ultraviolet and near-infrared photometry of FBQS J1408+3054, plotted as 
$F_\lambda$ versus rest frame wavelength in \AA.
The green points are the 2MASS data from 1999.
The blue spectrum is the Keck data from 2005. 
The red point is the GALEX NUV data from 2005.
The black points are the GALEX FUV and NUV data from 2007.
The error bars in the horizontal direction for the GALEX points
were obtained by taking the difference of the effective wavelength
and the limits of the passband in both directions and averaging. 
}
\label{fig:GALEXFig.eps}
\end{figure}

To add to our picture of \fbqs, we also searched for additional data
at other wavelengths (Figure~\ref{fig:GALEXFig.eps}).
We checked for sensitive X-ray coverage, but none is available.
\fbqs\ was detected by 2MASS \cite{2mass} in 1999 (MJD 51517).
With GALEX \cite{galex}, FBQS J1408+3054 was detected in the NUV filter in
2005 (MJD 53523.5) and in both the NUV and FUV filters in 2007 (MJD 54224.5).
The flux decrement between the NUV and FUV filters is consistent with the 
expected strong absorption in the Lyman transitions at the quasar redshift.
The FUV non-detection on MJD 53523.5 is consistent with the fainter NUV flux at the same epoch.
Both GALEX observations occurred after the MJD 53377 Keck spectrum which 
shows a strong reduction in \feii\ and \mgii\ absorption strength,
so we cannot determine if the absorption at observed NUV wavelengths has
weakened like that at observed optical wavelengths has.
The MJD 53523.5 NUV flux level is consistent with the MJD 53377 Keck spectrum,
and the MJD 54224.5 NUV flux level is consistent with 
a continued decrease in absorption.

\begin{table}
\begin{minipage}{80mm}
\caption{Spectroscopic Observations of FBQS 1408+3054}
\begin{tabular}{|c|c|c|c|c|c|}
\hline
Date       & MJD   & Source    & $\lambda/\Delta\lambda$
& $t$\footnote{Rest-frame days since the first spectroscopic epoch.}  
& $\Delta t$\footnote{Rest-frame days since the previous spectroscopic epoch.} \\ \hline
1995-05-23 & 49860 & KPNO 4m   & 1000 & 0    & ... \\
1996-01-11 & 50093 & Lick Kast &  835 & 126  & 126 \\
1997-04-10 & 50548 & KPNO 2.1m & 1250 & 372  & 246 \\
1997-05-16 & 50584 & Lick Kast &  835 & 392  & 20  \\
1997-06-10 & 50609 & Lick Kast &  835 & 405  & 13  \\
2000-03-09 & 51612 & Keck ESI  & 6750 & 948  & 543 \\
2005-01-07 & 53377 & Keck ESI  & 6750 & 1903 & 955 \\
2006-03-01 & 53795 & SDSS DR7  & 1850 & 2129 & 226 \\
2009-07-16 & 55028 & HET LRS   & 1130 & 2797 & 668 \\ \hline 
\end{tabular}
\end{minipage}
\end{table}


\begin{table*}
\begin{minipage}{100mm}
\caption{Observed and Synthesized Photometry of FBQS 1408+3054}
\begin{tabular}{|c|c|c|c|c|c|c|r|r|}
\hline
Date       & MJD   & Source  & Original & g  & r    & i & g$-$r & r$-$i \\ \hline
1950-04-18 & 33389 & POSS-I  & O=19.2 & 18.6 & ...  & ...   & ...  & ... \\
1950-04-18 & 33389 & POSS-I  & E=17.1 & ...  & 17.4 & ...   & 1.2  & ... \\
1982-05-21 & 45110 & Palomar & V=17.5 & 18.3 & ...  & ...   & ...  & ... \\
1992-04-08 & 48720 & POSS-II & J=19.0 & 18.8 & ...  & ...   & ...  & ... \\
1992-04-26 & 48738 & POSS-II & F=17.3 & ...  & 17.4 & ...   & 1.4  & ... \\
1995-02-23 & 49771 & POSS-II & N=17.0 & ...  & ...  & 17.4  & ...  & ... \\
1996-01-11 & 50093 & Lick$^a$ & ...   & ...  & ...  & ...   & 0.95 & $-$0.05 \\
1997-04-10 & 50548 & Lick$^a$ & ...   & ...  & ...  & ...   & 0.78 & ... \\
1997-05-16 & 50584 & Lick$^a$ & ...   & ...  & ...  & ...   & 1.28 & $-$0.01 \\
1997-06-10 & 50609 & Lick$^a$ & ...   & ...  & ...  & ...   & 1.15 & 0.05 \\
1999-12-05 & 51517 & 2MASS   & ...    & ...  & ...  & ...   & ...  & ...  \\
2000-03-09 & 51612 & Keck$^a$ & ...   & ...  & ...  & ...   & 0.95 & 0.09 \\
2004-04-13 & 53108 & SDSS   & g,r,i & 17.82 & 17.40 & 17.33 & 0.42 & 0.07 \\
2004-04-15 & 53110 & SDSS   & g,r,i & 17.84 & 17.40 & 17.34 & 0.44 & 0.06 \\
2005-01-07 & 53377 & Keck$^a$ & ...   & ...  & ...  & ...   & 0.48 & 0.07 \\
2005-06-02 & 53524 & GALEX   & ...    & ...  & ...  & ...   & ...  & ...  \\
2006-03-01 & 53795 & SDSS\footnote{Synthesized from spectroscopy; see text.} & ...   & 17.76 & 17.47 & 17.44 & 0.29 & 0.03 \\
2007-05-04 & 54225 & GALEX   & ...    & ...  & ...  & ...   & ...  & ... \\ \hline
\end{tabular}
\end{minipage}
\end{table*}


\section{Black Hole Mass Estimate}	\label{mbh}

To estimate the black hole mass of this quasar 
using the FWHM (Full Width at Half Maximum) of its H$\beta$ emission line, 
we must remove the neighboring \feii\ emission.

We subtracted \feii\ emission from the MJD 53377 Keck ESI spectrum
using two templates: Boroson \& Green (1992; BG92)
and V{\'e}ron-Cetty et al. (2004; VJV04),
both convolved with Gaussians of FWHM
spaced every 250 km~s$^{-1}$ from 1000 to 20,750 km~s$^{-1}$.
First, we subtracted a power-law
continuum fit made using observed-frame windows 7007-7079 \AA\ and 
10280-10351 \AA\ (rest-frame 3791-3830 \AA\ and 5562-5601 \AA).\footnote{This
fit lies above the observed spectrum around 4685~\AA,
suggesting that FBQS J1408+3054 is one of the rare class of
Balmer-line BAL quasars (Wang et al. 2008, and references therein).
\feii\ subtraction confirms that a significant H$\beta$ trough 
exists.  The trough is blueshifted by 11,000 km~s$^{-1}$, 
so it does not significantly affect the measurement of the H$\beta$ FWHM.
Although the H$\beta$ BAL appears to be deeper on MJD 51612 than on MJD 53377,
we cannot reliably claim a detection of variability.  
The relative flux calibration uncertainties are larger this close to the
long-wavelength end of the spectrum, and apparent variability in the 
\feii\ emission could affect the appearance of the H$\beta$ trough, 
but no H$\beta$ continuum fit or \feii\ subtraction is possible for the MJD 
51612 spectrum due to its limited wavelength coverage longward of H$\beta$.
}
Next, each velocity-broadened \feii\ template was scaled using a
progressively increasing normalization factor and the $\chi^2$ value between
the spectrum and each normalized template was measured using the rest-frame
region 4440-4647\,\AA.
The normalization factor with the lowest $\chi^2$ for any velocity FWHM was
chosen; the best-fit \feii\ FWHM was 3000 km~s$^{-1}$ in both cases.

The resulting \feii-subtracted
spectra were used to measure the H$\beta$ FWHM.
After subtraction of the VJV04 template,
we obtained FWHM=4100 km~s$^{-1}$.
After subtraction of the BG92 template,
we obtained FWHM=3800 km~s$^{-1}$.
 
To determine the black hole mass from H$\beta$, we use the formula
$M_{BH} = 5.5 {\rm FWHM}^2 R_{blr}/G$ from Onken et al. (2004),
where $G$ is the gravitational constant.
To determine $R_{blr}$ we use the formula in Bentz et al. (2009):
\begin{equation}
\log_{10} R_{blr} = -21.3^{+2.9}_{-2.8}
+0.519^{+0.063}_{-0.066}\log_{10}(\lambda L_{\lambda}(5100))
\end{equation}
where the H$\beta$ broad-line region radius $R_{blr}$ is in light-days,
$\lambda=5100$\,\AA\ rest-frame
and $L_\lambda(5100)$ is the monochromatic luminosity there.
We measure 
$f_\lambda(5100)=1.5\times 10^{-16}$~ergs~s$^{-1}$~cm$^{-2}$~\AA$^{-1}$,
which corresponds to 
$\lambda L_{\lambda}(5100)=2.66 \times 10^{45}$ ergs s$^{-1}$ in a Universe
with $H_0=70$ km~s$^{-1}$~Mpc$^{-1}$, $\Omega_M=0.3$ and $\Omega_\Lambda=0.7$.
These quantities yield $R_{blr}=190$ light-days 
(0.16 parsecs) or $4.9 \times 10^{12}$~km. 

We thus find 
$M_{BH}=3.4 \times 10^9$~$M_\odot$ using the VJV04 template
and $M_{BH}=2.9 \times 10^9$~$M_\odot$ using the BG92 template.
We adopt $M_{BH}=(3.15 \pm 0.35) \times 10^9$~$M_\odot$,
which yields a Schwarzschild radius of $R_{Sch}=9.3 \times 10^9$\,km,
so that $R_{blr} = 530 R_{Sch}$.
Using the mean bolometric correction of Richards et al. (2006), 
we estimate an unremarkable $L_{bol}/L_{edd} = 0.07 \pm 0.02$ for this object.

\section{Discussion} \label{discuss}

The trough variations in \fbqs\ are likely due to motions of 
absorbing gas transverse to our line of sight (e.g., Hamann et al. 2009).
Such motions are seen in numerical simulations of line-driven outflows, 
both two-dimensional (Proga, Stone \& Kallman 2000; Proga \& Kallman 2004)
and three-dimensional (Kurosawa \& Proga 2009).
In this section we first discuss why a changing ionization explanation
for the trough variations is unlikely.
Then we combine an estimate of the ultraviolet (UV) continuum source size
in this object with the timescale over which the absorption disappears
to estimate the transverse velocity of the absorbing gas.
We then outline a simple BAL outflow model with which 
we constrain the distance of the absorbing gas from the black hole
and the time since the gas was launched as part of the outflow.
We close by discussing the implications of our results 
for the connections between BAL subtypes,
for the average episodic lifetimes of BAL outflows 
and for the link between FR~II radio morphology and BAL quasars.

\subsection{Was the Fe\,{\sc ii} Ionized Away?} \label{ion}

One reason
it is unlikely that the changes in the BAL troughs of \fbqs\ are due to a
increased ionization of the absorbing gas is because
the near-UV continuum flux of this quasar did not increase
significantly while the absorption troughs weakened dramatically.
The 1997 spectrum presented in Becker et al. (2000) and the 2006
SDSS spectrum show the same peak flux density near \mgii\ to within 10\%,
and SDSS photometry of \fbqs\ taken on MJD 53110 yields $gri$ magnitudes
within $\pm$10\% of $gri$ magnitudes synthesized from SDSS spectroscopy
taken on MJD 53795.

Of course, ionization is caused not by near-ultraviolet photons but by extreme
UV and soft X-ray photons with $\lambda \leq 912$\,\AA, and along BAL quasar
sightlines such photons are thought to be absorbed by a layer of shielding gas
(e.g., Gallagher et al. 2006).  It is possible that a decrease in the amount
of shielding gas along our line of sight could have increased the ionizing flux
without increasing the near-UV flux that we have observed, resulting in the
ionization of the \feii\ absorber to \feiii.  Increasing ionizing flux would
have expanded the \feii\ ionization front along our line of sight outwards.
The timescale for that expansion is the timescale for the decrease in the
shielding gas column or the recombination timescale in the \feii\ absorber,
whichever is longer.
The recombination timescale is $(\alpha n_e)^{-1}$,
where $\alpha$=1.02$\times$10$^{-12}$ cm$^{-3}$ s$^{-1}$ is the
total recombination coefficient of \feii\ at $T$=10$^4$ K (Woods et al. 1981)
and $n_e$ is the \feii\ absorber's electron density.

We cannot definitively rule out a change in ionizing flux as the explanation
for the disappearance of \feii\ absorption in \fbqs\ because we lack X-ray data
bracketing the disappearance epoch.
We do not believe such an explanation is likely, however, because
no evidence for ionization-dependent absorption variations has been
found for the general population of BAL quasars (Gibson et al. 2008).
Nonetheless, if a decreasing shielding gas column was responsible for 
\feii\ vanishing from the spectrum of \fbqs, we can set a lower limit on $n_e$
in the \feii\ absorber of between
$ 6\times 10^3 $ cm$^{-3}$ and $ 5\times 10^4 $ cm$^{-3}$ corresponding to the
decrease occurring on the maximum or minimum observed timescales, respectively.

\subsection{Constraints on the BAL Structure Size and Transverse Velocity} \label{sizevt}

We refer to the gas that absorbs the quasar's light in \mgii\ and
\feii\ as the BAL structure.  Our conclusions about its size and 
location hold regardless of whether the structure is a single large cloud,
part of a continuous flow tube (e.g., Arav et al. 1999),
or a collection of smaller subunits (e.g., Hall et al. 2007).
The column density of the absorption may vary over the structure even in the
first two cases above (de Kool et al. 2002; Arav et al. 2005; Sabra \& Hamann
2005), but for convenience we assume the structure has a sharp edge.
Note that the entire BAL outflow can be much larger in size than the 
BAL structure we observe along our sightline (including extending to
greater distances from the quasar) and can have a longer lifetime as well.

Given the size of the continuum source,
the timescale over which absorption changes from strong to weak 
(or vice versa) constrains the velocity and location of the BAL structure. 
Furthermore, the length of time the quasar was seen as a BAL quasar
constrains the size of the BAL structure in one transverse direction.
For our discussion, we take the change in the spectrum of \fbqs\ that began
between 2000 and 2005 to be a change from strong to weak absorption.

The minimum timescale over which the \feii\ troughs disappeared is the time
between the 2005 and 2006 spectra: $\Delta t_{min}$=226 rest-frame days.
The maximum allowed timescale is the time 
between the 2000 and 2009 spectra: $\Delta t_{max}$=1849 rest-frame days.
We adopt an intermediate timescale of $\Delta t=946$ rest-frame days
as the time required for the \mgii+\feii\ rest-frame equivalent width
to decrease from its level in
2000 to its level in 2009 at the rate of change observed between 2005 and 2006.

This value of $\Delta t=946$ rest-frame days
is an estimate of the crossing time of the quasar 
emission region by the {\em trailing edge} of the BAL structure.\footnote{We
have assumed a sharp-edged structure, but the results apply generally.  Because
the \feii\ absorption goes from its maximum to zero in $\simeq$946 rest-frame
days, a diffuse structure in which $N_H$ gradually changes at the edges would
have to move across the face of \fbqs\ faster than a sharp-edged structure.
Therefore, the minimum transverse velocity derived for a sharp-edged structure
is a lower limit to the velocity of any structure.}
As the quasar was seen as an FeLoBAL quasar of roughly constant $W_r$ for 
at least 948 rest-frame days prior to the 2000 epoch, the time required for
the {\em entire} BAL structure observed in \fbqs\ since its discovery 
to cross the quasar emission region was at least $t_{cross}=2\Delta t$.  
In other words, the entire BAL structure took at least twice as long to cross
the quasar emission line region as the edge of the BAL structure 
did.\footnote{The lack of 
strong variability in the historical photometry of \fbqs\ discussed
in \S~\ref{Multi} suggests that the BAL structure actually had a crossing time
of $\gtrsim$27 rest-frame years, $\sim$10 times longer than the crossing time
of its edge.  However, as we lack spectroscopy or regularly sampled photometry
in pre-discovery epochs, we quote the shorter timescale to be conservative.}
Twice the 2700 \AA\ continuum emission region size $D_{2700}$ is thus an
approximate lower limit on the transverse size of the BAL structure in the
direction of its projected motion on the sky.
The maximum depth of the \feii\ absorption reveals that 
the BAL structure covered at least $\sim$75\% of the
region that produces the continuum emission at 2700 \AA\ (and possibly part
of the \feii-emitting region, but not all; see below).
Because this covering fraction was roughly constant for several rest-frame
years, we take it to be the extent of the BAL structure in the 
direction on the sky orthogonal to the structure's motion on the sky.

We use a Shakura \& Sunyaev (1973)
thin accretion disk model to estimate $D_{2700}$, 
the diameter of the disk within which 90\% of the 2700~\AA\ continuum 
is emitted, for a $M_{BH}=3.15 \times 10^9$ $M_\odot$ black hole.
We find $D_{2700}=4.28 \times 10^{11}$ km (46$R_{Sch}$).
The projected dimensions of the BAL structure are therefore
$\gtrsim 3.2 \times 10^{11}$\,km or 0.01 pc wisde
by $\gtrsim 8.6 \times 10^{11}$\,km or 0.028 pc
(and likely $\gtrsim 8.6 \times 10^{12}$\,km or 0.28 pc)
long in its projected direction of motion.

Using $\Delta t$ and $D_{2700}$, we can constrain the
velocity of the BAL structure perpendicular to our line of sight: 
$v_{\perp} = D_{2700}/\Delta t = 5200_{-2600}^{+16800}$ km~s$^{-1}$,
where the range comes from
the constraints on $\Delta t$.
The uncertainty on the size of the region crossed by the flow can change
these limits somewhat (see the end of the next section).
Nonetheless, the transverse velocity is constrained to be approximately
2600~km~s$^{-1}$ $\leq$ $v_{\perp}$ $\leq$ 22,000~km~s$^{-1}$);
for comparison, the
line-of-sight velocity of peak absorption is a blueshift of 12,000 km~s$^{-1}$.

We can rule out the possibility that the BAL structure 
completely crossed the larger \feii-emitting region during the time span 
$\Delta t$.  If we use $2R_{\rm FeII}$ instead of $D_{2700}$ as the 
size of the region crossed by the BAL structure, $v_{\perp}$ will increase.  
As the \feii\ FWHM in this object is smaller than the H$\beta$ FWHM,
\feii\ emission is likely to come from a larger average radius than H$\beta$.
Assuming $R\propto{\rm FWHM}^{-2}$, we obtain
$R_{\rm FeII} = 1.74R_{blr} = 8.5 \times 10^{12}$~km = 910$R_{Sch}$.
In that case, we find 
$v_{t,FeII} = 2R_{\rm FeII}/\Delta t = 206,000$ km~s$^{-1}$. 
Such a high velocity implies a small distance from the quasar, which is not
self-consistent.
A BAL structure located at any distance $d_{BAL} < R_{\rm FeII}$ would not
be able to cover the full extent of the \feii\ emission region.

\subsection{A Model for Constraining the BAL Structure Location}

When the radial velocity of absorbing gas around a quasar is not known, 
its distance from the black hole is usually estimated by assuming it is in 
a circular orbit (e.g., Risaliti et al. 2007).  
However, in our case the radial velocity of the BAL gas is 
12,000 km~s$^{-1}$ and we cannot self-consistently assume a circular orbit.
To estimate the distance of the BAL structure from the black hole, $d_{BAL}$,
some assumptions about the velocity field of the BAL wind must be made
(see, e.g., \S~6.2 of Misawa et al. 2005).  

Given the large blueshift of the BAL structure in \fbqs, the
outflow was likely launched at a radius where the escape velocity is comparable
to the observed blueshift (e.g., Murray et al. 1995; Proga, Stone \& Drew 1998; 
Proga 1999; Everett 2005; Zhang \& Thompson 2010).  A natural source for such gas
is a wind from the accretion disk around the quasar,
and that is the scenario we adopt.

\begin{figure*}
\includegraphics[width=84mm]{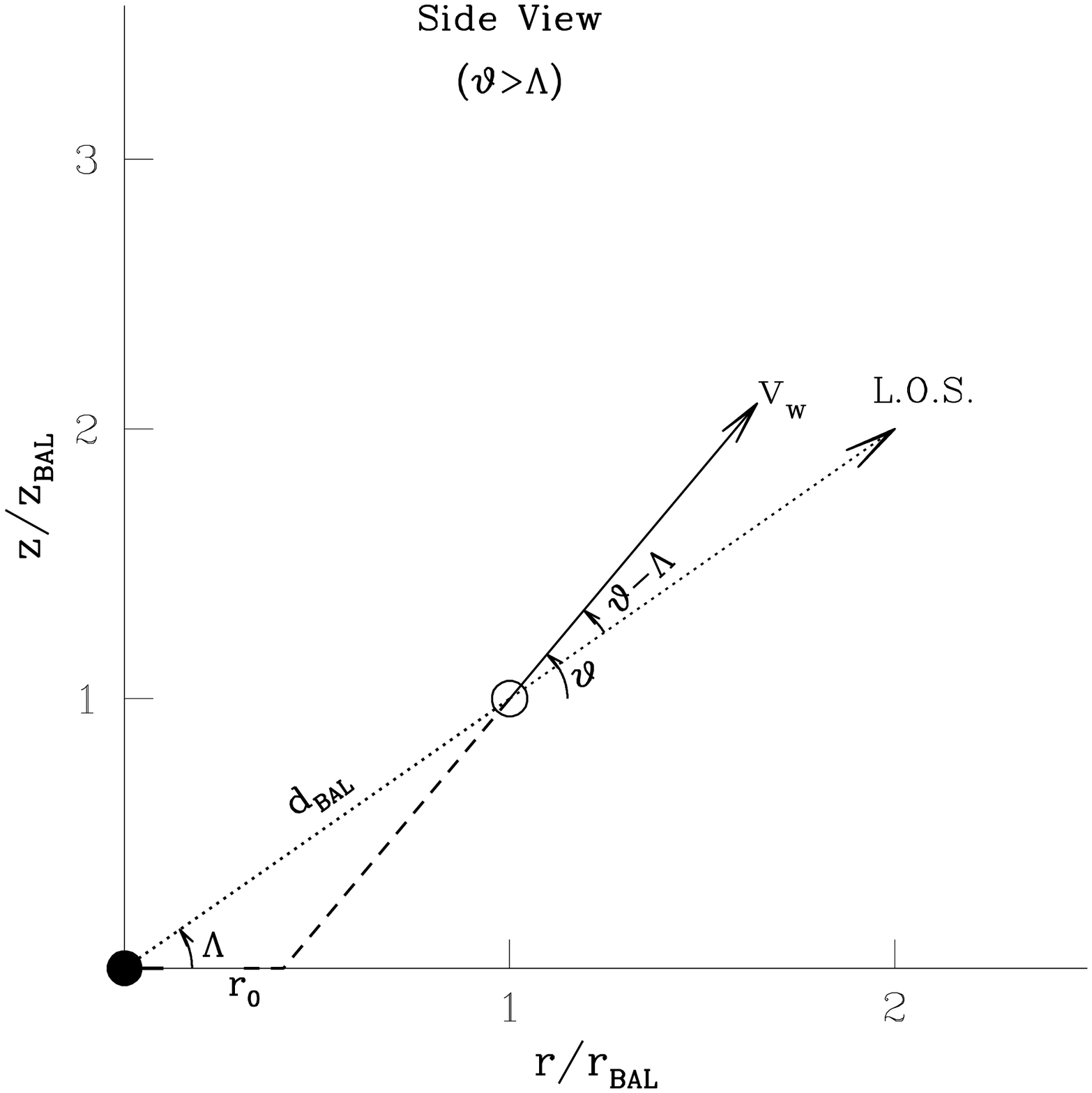}
\includegraphics[width=84mm]{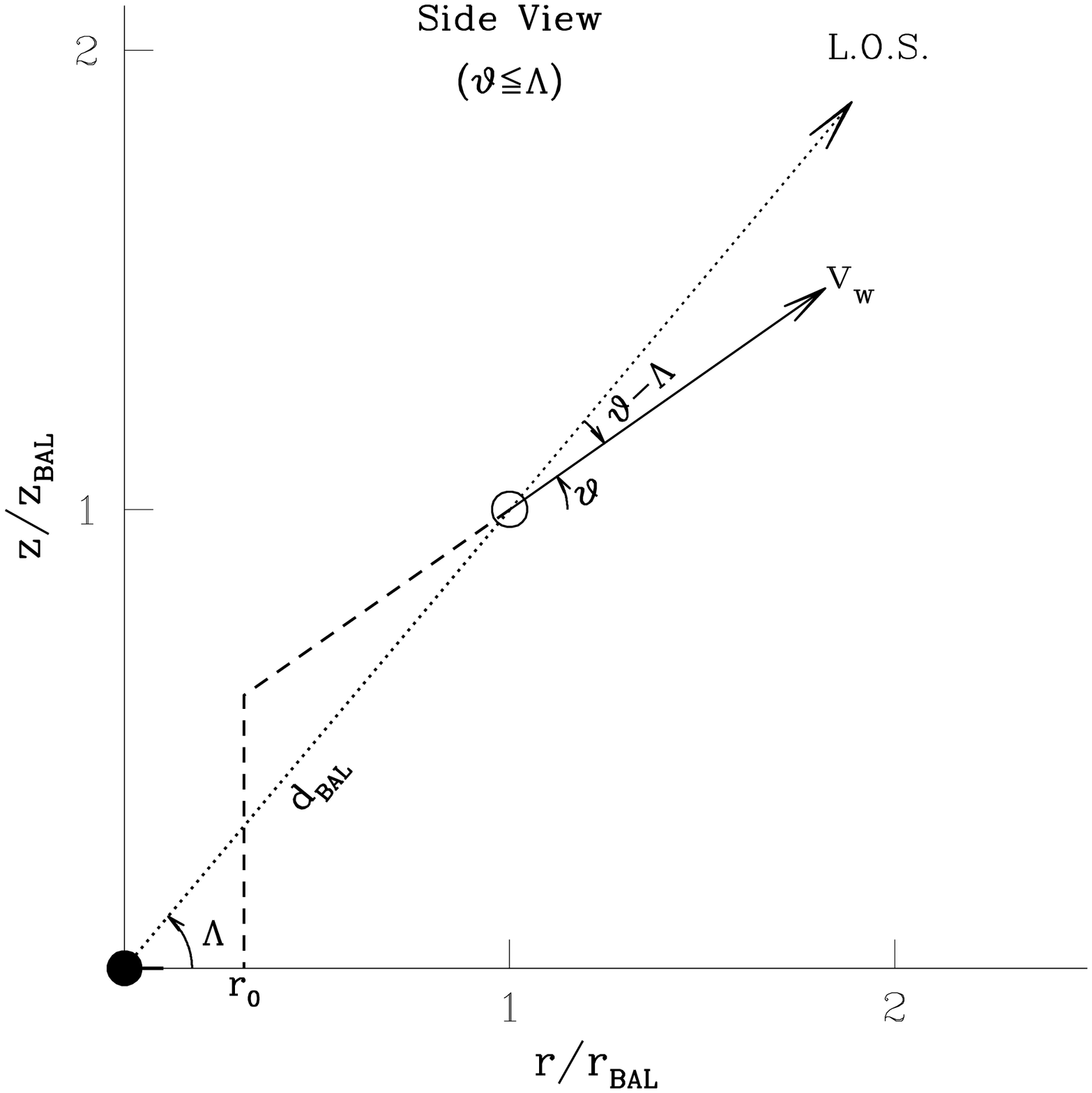}
\caption{Side view of a BAL structure (open circle)
obscuring the our line of sight (L.O.S.) to a black hole and accretion disk
(filled circle at the origin and thick horizontal line segment).
Our line of sight is at an angle $\Lambda$ above the disk. 
The BAL structure at location ($r_{BAL},z_{BAL}$) moves 
at velocity $V_w$ at angle $\vartheta$ above the disk in the $r-z$ plane
and at azimuthal velocity $V_\phi$ normal to the $r-z$ plane.
The dashed line segments show a plausible upper limit on the projected 
distance the BAL wind has traveled from its launch radius.  Two geometries
that yield equal and opposite $\vartheta-\Lambda$ are shown:
on the left, $\vartheta=50^\circ$ and $\Lambda=35^\circ$;
on the right, $\vartheta=35^\circ$ and $\Lambda=50^\circ$.}
\label{sideview}
\end{figure*}

We assume that our line of sight to the quasar lies at a latitude $\Lambda$
above the accretion disk and that the accretion disk has azimuthal symmetry
(Figure \ref{sideview}).
The BAL wind crosses our line of sight to the continuum emission region
at a distance $d_{BAL}$ from the black hole, at cylindrical coordinates
$r_{BAL}=d_{BAL}\cos\Lambda$ and $z_{BAL}=d_{BAL}\sin\Lambda$.
We parametrize the velocity field of the BAL wind at this point
as an azimuthal velocity $V_\phi$ and a poloidal velocity $V_w$ 
oriented at an angle $\vartheta$ above the accretion disk.\footnote{The 
magnitude of the
three-dimensional velocity $\mathsf v$ of the BAL wind at this point is
given by ${\mathsf v}=\sqrt{V_\phi^2+V_w^2}=\sqrt{v_\perp^2+v_\parallel^2}$.}
The BAL velocity components parallel and perpendicular to our line of sight are
\begin{equation}\label{parallel}
v_\parallel = V_w \cos(\vartheta-\Lambda) \end{equation}
\begin{equation}\label{perpendicular}
v_\perp = \sqrt{V_\phi^2 + V_w^2\sin^2(\vartheta-\Lambda)}. \end{equation}
We know $v_\parallel$ from spectroscopy,
and the BAL variability constrains $v_{\perp,min} \leq v_\perp \leq v_{\perp,max}$.

\begin{figure*}
\includegraphics[width=84mm]{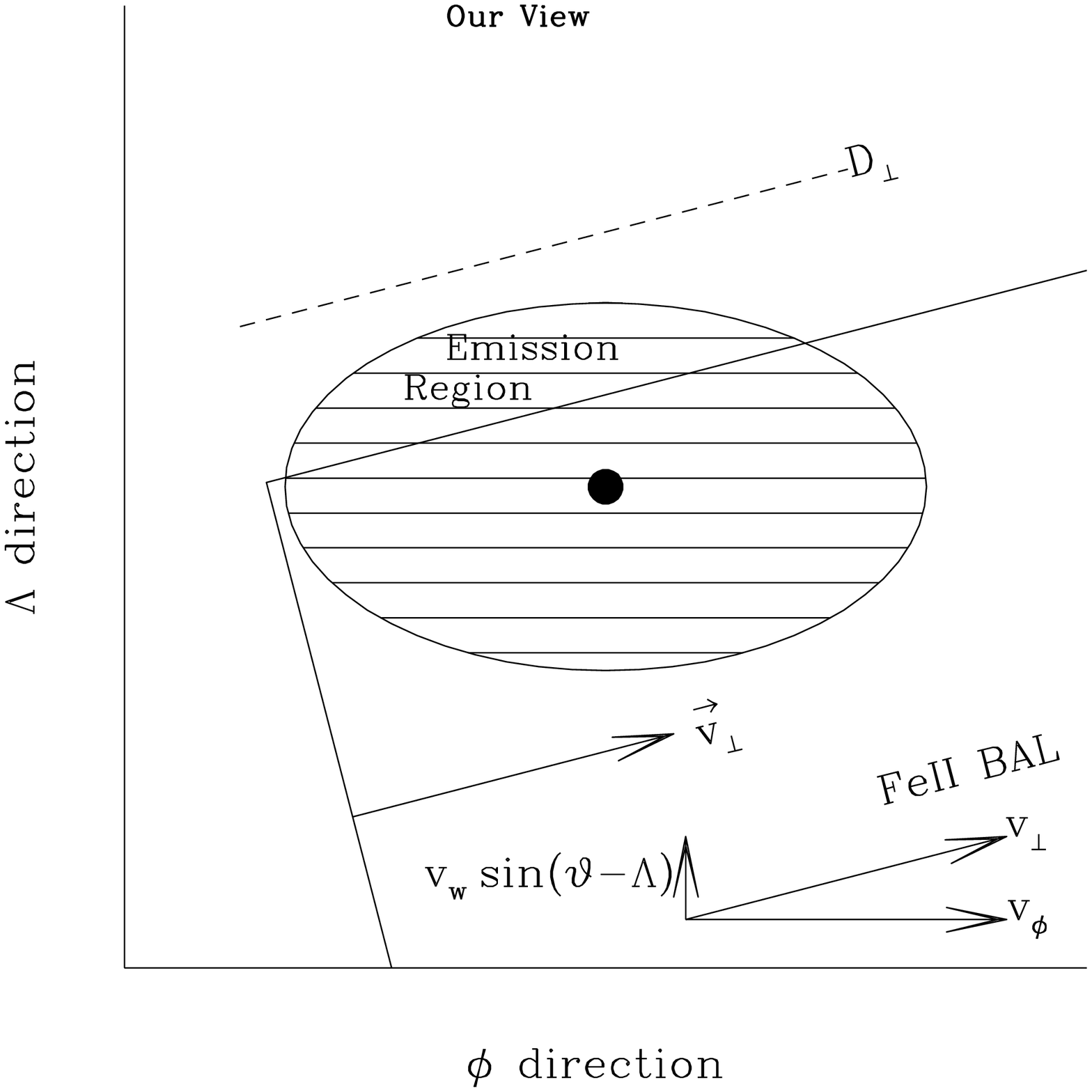}
\includegraphics[width=84mm]{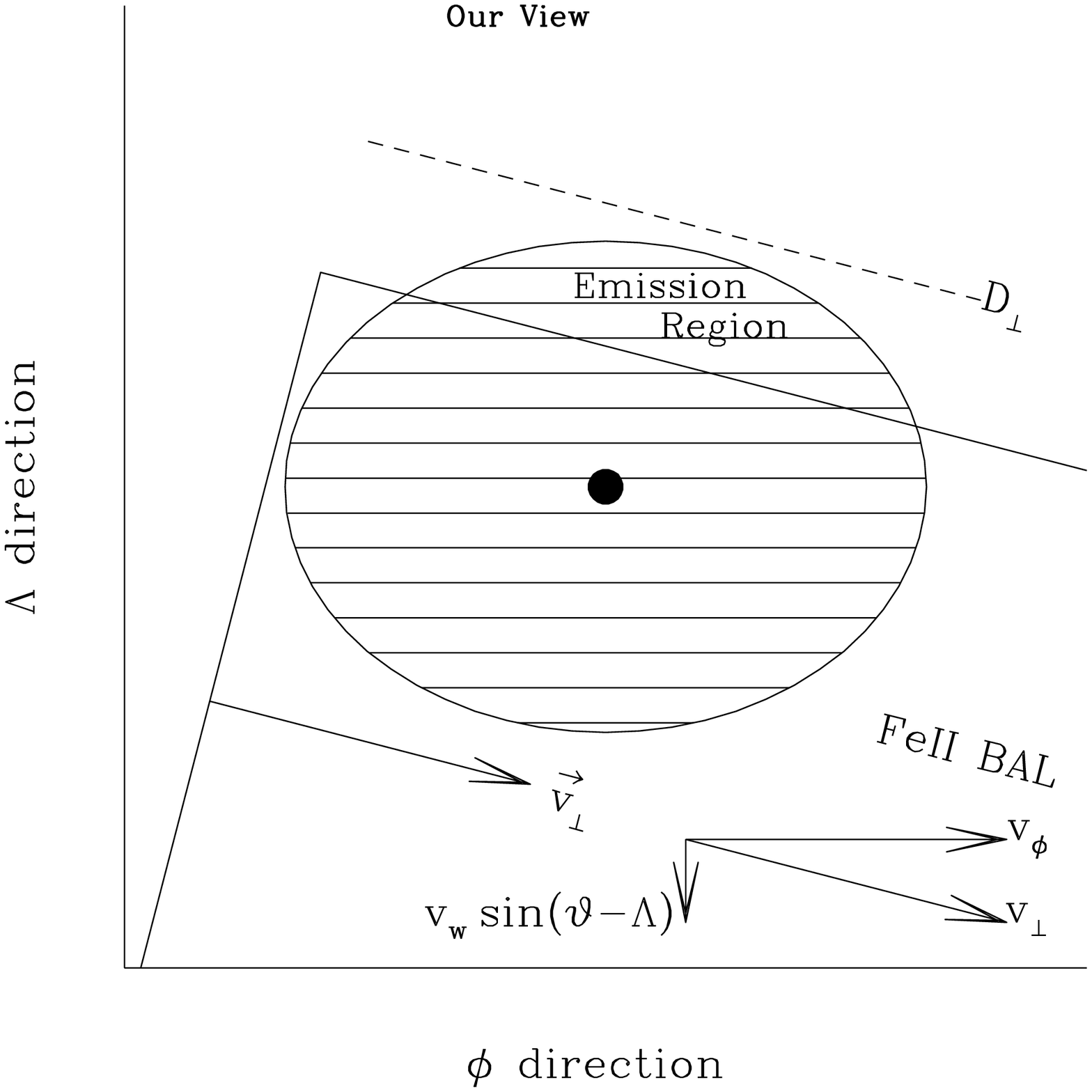}
\caption{View from along our line of sight of a BAL structure (large rectangle
extending off the plot) partially obscuring the continuum emission region
around the black hole.
The continuum emission region, indicated here with horizontal lines,
is intrinsically a circular disk but is projected to an ellipse.
The BAL structure moves with velocity $\vec{v}_\perp$, 
which can be broken down into two components: the azimuthal velocity $V_\phi$
and the component of the wind velocity in the plane of the sky 
$V_w\sin(\vartheta-\Lambda)$.
The same two geometries as in the previous figure are shown:
on the left, $\vartheta=50^\circ$ and $\Lambda=35^\circ$;
on the right, $\vartheta=35^\circ$ and $\Lambda=50^\circ$,
in which the emission region appears more circular.
The width of the continuum emission region projected along the $\vec{v}_\perp$
direction, $D_\perp$, is indicated by the dashed line segment; $D_\perp$ is
slightly larger in the geometry shown on the right.
Although the values of $\vartheta$ and $\Lambda$ are unknown in \fbqs, these
figures are otherwise representative of \fbqs, where we observed the
disappearance from along our line of sight of an \feii\ BAL structure with 
a minimimum size larger along $\vec{v}_\perp$ than perpendicular to it 
and which even at maximum absorption likely covered only $\sim 75$\% 
of the continuum source.
}
\label{ourview}
\end{figure*}

To link these velocities to $d_{BAL}$, we first assume that the BAL outflow
started out in a circular orbit at radius $r_0$ in the accretion disk 
and has not swept up a significant amount of mass since then.
We assume conservation of angular momentum in the $\phi$ direction
(neglecting momentum carried by magnetic fields) so:
\begin{equation}\label{vphi}
V_\phi =  \sqrt{G M_{BH} r_0} / r_{BAL}
\end{equation}
where $r_{BAL}$ is the current distance
of the BAL structure from the axis of symmetry of the accretion disk.

Next, we use the results of Murray et al. (1995, fig. 3) 
and Murray \& Chiang (1998, equation 2)
that the final poloidal
velocity of a radiatively driven outflow launched from radius $r_0$
is $V_{\infty}=(3.6\pm 1.1)\sqrt{GM_{BH}/r_0}$ and that $V_{\infty}$ is reached
within a radial distance of a few times $r_0$.
Therefore, we assume
\begin{equation}\label{vinfty}
V_w = V_{\infty} = (3.6\pm 1.1)\sqrt{GM_{BH}/r_0} \end{equation}
and write
\begin{equation}\label{para}
v_\parallel = (3.6\pm 1.1)\sqrt{GM_{BH}\over r_0} \cos(\vartheta-\Lambda)
\end{equation}
\begin{equation}\label{perp}
v_\perp=\sqrt{{GM_{BH}r_0\over r_{BAL}^2}+(3.6\pm 1.1)^2{GM_{BH}\over r_0}\sin^2(\vartheta-\Lambda)}.
\end{equation}
These equations express $v_\parallel$ and $v_\perp$ in terms of 
physical parameters in our outflow model, in principle enabling 
us to constrain those parameters and place limits on
$d_{BAL}=r_{BAL}/\cos\Lambda$.

In the case of known $\Lambda$ and $\vartheta$, 
$r_0$ can be obtained from equation \ref{para}.
A lower limit on $v_\perp$ then yields an upper limit on $r_{BAL}$ via
equation \ref{perp}, and an upper limit on $d_{BAL}$ follows. 
Similarly, a lower limit on $d_{BAL}$ follows from an upper limit on $v_\perp$.

Note that if $\Lambda > \vartheta$, the streamlines must have bent
to reach the location $(r_{BAL},z_{BAL})$ with velocity $V_w$
at angle $\vartheta$.  
The streamlines therefore must also have crossed our line of sight
at another position closer to the black hole (Figure \ref{sideview},
right-hand panel), possibly producing another BAL trough in the spectrum.  
Another trough is guaranteed to be produced only if the BAL outflow
is continuous back to the launch radius and extends over all azimuthal angles.

We make one final refinement to the above analysis.  We have assumed that
continuum emission region is intrinsically circular with diameter $D_{2700}$, 
projected on the sky to an ellipse of major axis $D_{2700}$.
The transverse velocity limits 
were computed assuming a distance $D_{2700}$ was crossed by the BAL flow,
but in fact the trailing edge of the BAL flow 
almost always crosses a distance $D_\perp < D_{2700}$ (Figure \ref{ourview}). 
$D_\perp$ is defined as the distance spanned by the projection of the ellipse
onto an axis parallel to $\vec{v}_\perp$;
only when $\Lambda=\vartheta$ does $D_\perp=D_{2700}$.
For a given $\vartheta$ and $\Lambda$, once we calculate $r_0$ from equation
\ref{para} we calculate the direction of $\vec{v}_\perp$ on the sky and the
distance $D_\perp$ along that direction, then recalculate 
\begin{equation}\label{vperpminmax}
v_{\perp,min}={D_\perp \over \Delta t_{max}};~~
v_{\perp,max}={D_\perp \over \Delta t_{min}}
\end{equation}
for comparison to equation \ref{perp} and iterate to convergence.

\subsection{Constraints on the BAL Structure Location}

Unfortunately,
we only have statistical constraints on $\Lambda$ and $\vartheta$.
We must therefore consider what constraints we can put on $r_0$ and $r_{BAL}$
given reasonable distributions of our viewing angle $\Lambda$ 
and the wind velocity's poloidal angle above the disk plane $\vartheta$.

We assume $0^\circ < \Lambda < 73^\circ$ 
from the size of this FR~II quasar's radio lobes
relative to the maximum size known for such lobes (Gregg et al. 2006).
(Even without identifying this quasar as an FR~II quasar,
we can assume $\Lambda \leq 80^\circ$ because Hovatta et al. (2009) 
find that objects viewed at $\Lambda > 80^\circ$ are seen as blazars.)
We assume $5^\circ \leq \vartheta \leq 85^\circ$ for the outflow angle
because we have no real constraints on this parameter.
We sample both angles using uniform distributions in solid angle.

We compute lower and upper limits on $d_{BAL}$ for each combination of
$\Lambda$ and $\vartheta$ spanning the above limits, and take the averages.
Combinations for which equation \ref{perp} yields the square root of a negative
number as an upper limit on $v_\perp$ are ruled out in this model and
do not place any constraints on $d_{BAL}$.  Also ruled out are combinations
that yield
$r_{BAL,max} < r_0$ for $\vartheta\leq \Lambda$ or
$r_{BAL,max} < r_0/(1-\tan\Lambda/\tan\vartheta)$ for $\vartheta>\Lambda$.

There are two sources of uncertainty in the lower and upper limits we find.
The first is the scatter in results between different $\vartheta$,$\Lambda$
combinations.  This uncertainty can be reduced with better knowledge of
quasar viewing angle distributions\footnote{For example, we have assumed a flat
probability distribution of the viewing angle $\Lambda$ in \fbqs, 
0$^\circ$$<$$\Lambda$$<$73$^\circ$.  
A more realistic probability distribution could be
constructed by comparing the observed separation between its radio lobes to an
inferred distribution of true lobe separations, or by using other methods
of inferring radio source orientations (Richards et al. 2001).} 
or FeLoBAL wind outflow angles.
However, even without such knowledge, the ratio of upper to lower limits
on the BAL distance is well constrained; the values of such limits vary
considerably with $\vartheta$ and $\Lambda$, but their ratio does not.
The second is the uncertainty in inferring $r_0$ from the factor of 3.6$\pm$1.1
in equation \ref{vinfty}.  The larger this factor, the farther out the wind
can be launched and still end up with the same observed velocity, for any
$\vartheta$,$\Lambda$ combination.  Thus, the BAL distance increases with
this factor.  This uncertainty will remain unless
studies of disk winds can refine the distribution
of their asymptotic wind velocities.

Including the above uncertainties,
we find average lower and upper limits on $d_{BAL}$ of
$5800_{-3900}^{+9400}\leq d_{BAL}/R_{sch} \leq 46000_{-42000}^{+129000}$
(90\% confidence ranges).
While these limits are each uncertain by more than an order of magnitude,
the ratio of upper to lower limits has a much more well defined value of
$\sim$7.5$\pm$1.5.

The large uncertainties are in part inherent to our limited knowledge of
the orientation of the outflow, but are also influenced by the fact that
the variability timescale (and thus the transverse velocity) in this object
is only constrained to within a factor of eight.
Better limits on the variability timescale 
would yield tighter constraints on the BAL location and restrict
the $\vartheta,\Lambda$ combinations which could match the observations.

For reference,
in Table 3 we list the average lower and upper limits on the distance of
the BAL structure in \fbqs\ as a function of the assumed ratio between
the BAL's final and initial poloidal velocities.
The 90\% confidence limits
in the Table come only from the scatter in the limits
for different $\vartheta,\Lambda$ combinations.

\begin{table}
\begin{minipage}{80mm}
\caption{BAL Distance Limits, Averaged Over $\vartheta$ and $\Lambda$}
\begin{tabular}{|c|c|c|c|}
\hline
${V_\infty\over\sqrt{GM_{BH}/r_0}}$ & $\langle{d_{BAL,min}\over R_{sch}}\rangle$ & $\langle{d_{BAL,max}\over R_{sch}}\rangle$ & $\langle$Ratio$\rangle$ \\ \hline
2.5 & $2900_{-1000}^{+1800}$ & $29000_{-25000}^{+48000}$ & 9.1$\pm$0.2 \\
3.6 & $5800_{-1900}^{+3100}$ & $46000_{-40000}^{+73000}$ &  7.0$\pm$0.2 \\
4.7 & $9900_{-3400}^{+5300}$ & $71000_{-62000}^{+104000}$ &  6.3$\pm$0.2 \\
\hline
\end{tabular}
\end{minipage}
\end{table}

These limits on $d_{BAL}$ place the BAL structure in \fbqs\ between
11 and 88 times farther from the quasar than the H$\beta$ broad-line region.
Physically, these limits correspond to a distance between 1.7 and 14~pc
from the black hole.  

Combinations of $\vartheta$ and $\Lambda$ that match the observations
have an average value of $\vartheta-\Lambda \simeq -16^{\circ}$, 
with a range of $-60^\circ < \vartheta-\Lambda < 41^\circ$
out of the allowed range of $-68^\circ < \vartheta-\Lambda < 85^\circ$.
The angle $\vartheta-\Lambda$ and the magnitudes of $V_w$ and $V_\phi$
constrain the angle $\psi$ between the line of sight and the
velocity vector of the wind.
For combinations of $\vartheta$ and $\Lambda$ that match the observations
we find an average $\psi\simeq 24^\circ \pm 2^\circ$,
with a range of $8.9^\circ < \psi < 60^\circ$
out of the allowed range of $1.2^\circ < \psi < 61.4^\circ$.

\subsection{Constraints on the BAL Structure Age} \label{tage}

The age of an observed BAL structure (how long it has existed)
is a lower limit on its lifetime (how long it will exist)
and on the lifetime of the entire BAL outflow.
A BAL structure's age can be constrained by considering how long
gas in the structure took to reach its current location after being launched.
This constraint applies regardless of whether the structure 
is the result of a single ejection event or part of a continuous flow.

To constrain the age of the BAL structure,
we trace the wind at angle $\vartheta$ from $(r_{BAL},z_{BAL})$ back to
$r=r_0$ or $z=0$, whichever comes first, and then back to $(r_0,0)$;
see Figure \ref{sideview}.
Such a path yields the longest plausible distance $s$ for the BAL structure
to have covered to reach its current position.  The formulae for $s$ are:
\begin{equation}
\tan\vartheta \geq {r_{BAL}\tan\Lambda\over r_{BAL}-r_0}:~
s=r_{BAL}{\tan\Lambda \over \sin\vartheta}(1-\cos\vartheta) + r_{BAL} -r_0
\end{equation}
\begin{equation}
\tan\vartheta \leq {r_{BAL}\tan\Lambda\over r_{BAL}-r_0}:~
s={r_{BAL}-r_0\over \cos\vartheta}(1-\sin\vartheta) +r_{BAL}\tan\Lambda.
\end{equation}
The age of the BAL structure is then simply $t_{age}=s/V_w$.
This value will be a slight underestimate
because the wind was launched with $v<V_w$,
but the underestimation will be small for $s\gg r_0$.

For combinations of $\vartheta$ and $\Lambda$ that match the observations of
\fbqs, we find average BAL structure ages of
$t_{age} = 90_{-70}^{+90}$ years for $d_{BAL} = 5800~R_{sch}$
and $t_{age} = 1100_{-1000}^{+1800}$ years for $d_{BAL} = 46000~R_{sch}$.
These lower limits on the BAL outflow lifetime in \fbqs\ are consistent with
other, independent determinations (see \S~\ref{Multi} and \S~\ref{tB}).

\subsection{Implications for Links Between BAL Subtypes}

Aside from their defining broad absorption at UV wavelengths
and accompanying X-ray absorption (e.g., Gallagher et al. 2006),
HiBAL quasars as a population do not appear distinct from non-BAL quasars
in terms of their multiwavelength properties
(Lewis et al. 2003; 
Willott et al. 2003; 
Priddey et al. 2007; 
Gallagher et al. 2007; 
Shen et al. 2008). 
The only exception is the greater reddening of HiBAL quasars
relative to non-BAL quasars (Brotherton et al. 2001; 
Reichard et al. 2003; Trump et al. 2006; Gibson et al. 2009).

Numerous studies have suggested that LoBAL and FeLoBAL quasars may have some
optical and mid-infrared (mid-IR) spectral properties distinct from HiBAL and 
non-BAL quasars, such as weaker [O\,{\sc iii}] emission, stronger 
\feii\ emission and continuum polarization in the optical and ultraviolet,
and higher mid-IR luminosities and different mid-IR spectra
(for LoBALs and the low-redshift FeLoBAL quasar Mrk 231, see
Low et al. 1989, 
Weymann et al. 1991, 
Boroson \& Meyers 1992 
and Canalizo \& Stockton 2001; 
for both LoBALs and FeLoBALs, see 
Schmidt \& Hines 1999 and 
Urrutia et al. 2009; 
for FeLoBALs, see
Farrah et al. 2007, 2010).
On the other hand, some studies have found little evidence that LoBAL 
or FeLoBAL quasars have properties significantly different from those 
of HiBAL quasars (for polarization properties, Ogle et al. 1999;
for mid-IR properties,
Lewis et al. 2003; Willott et al. 2003; Gallagher et al. 2007).

The transformation of \fbqs\ from an FeLoBAL to a LoBAL quasar 
along our line of sight casts these studies in a somewhat different light.
Suppose that at least some
FeLoBAL quasars have properties distinct from HiBAL and LoBAL
quasars in some wavelength range where the timescale for variability is 
longer than in the optical (e.g., in the mid-IR; Farrah et al. 2007, 2010).
Then the transformation of \fbqs\ from an FeLoBAL to a LoBAL requires that some,
but not necessarily all, LoBAL quasars must have the same distinct properties.
In this scenario, the reason why some but not all LoBAL quasars have unusually
high mid-IR luminosities as compared to HiBAL quasars has two underlying causes.
First, at least some FeLoBAL quasars have such high mid-IR luminosities.
Second, some objects that are seen as LoBAL quasars along our line of sight
could be classified as FeLoBAL quasars when seen along some lines of sight,
while some LoBAL quasars would not be seen as FeLoBAL quasars along any 
line of sight.

If FeLoBAL quasars as a population do have unusual multiwavelength properties,
they are unlikely to be ``normal'' BAL quasars seen at specific viewing angles.
They could instead be quasars emerging from a shroud of gas and dust 
(Voit et al. 1993), for example as transition objects between 
ultraluminous infrared galaxies (ULIRGs) and quasars (Farrah et al. 2010),
or quasars with a specific disk wind geometry (Richards 2006).
Our study of \fbqs\ does not provide support for the transition object scenario
because
the FeLoBAL gas in this object moves at very high velocities and is found only
1.7 to 14~pc from the black hole.  If FeLoBALs are to be transition objects
between ULIRGs and quasars, they must drive out gas on kiloparsec and not
just parsec scales.  That said, our study does not rule out this scenario
either.  Distance constraints on more FeLoBAL outflows would be one way to
determine how common transition objects are among FeLoBAL quasars.
Another possible test of the transition object scenario is to determine if 
FeLoBAL quasar host galaxies differ from the host galaxies of other quasars,
due to more recent interactions or mergers.

However, such differences are not necessarily required in the original
`clearing shroud' scenario of Voit et al. (1993).  That work postulated
that LoBAL quasars could be quasars emerging from parsec-scale shrouds
of dusty gas, and our observations are entirely consistent with such a 
scenario.  The existence of dusty gas on such small scales need not
depend on the kiloparsec-scale structure of the quasar host galaxy, 
so FeLoBAL quasar host galaxies could be unremarkable in this model.

In the alternate scenario where FeLoBALs are quasars with specific disk wind
geometries, they must be different from other quasars in at least one parameter
driving accretion disk structure.  
For example, while the distribution of Eddington ratios
is known not to differ between non-BAL and HiBAL quasars (Ganguly et al. 2007),
the FeLoBAL quasar Eddington ratio distribution has not been studied; 
however, \fbqs\ has an unremarkable $L_{bol}/L_{Edd}=0.07$ (\S~\ref{mbh}).

It is known that greater {\em average} reddening is found 
among ever rarer BAL subtypes.  That is, as mentioned above 
HiBAL quasars are on average more reddened than non-BAL quasars,
LoBAL quasars are more reddened than HiBAL quasars (Sprayberry \& Foltz 1992),
and FeLoBAL quasars are more reddened than LoBAL quasars (Reichard et al. 2003).
If variability of the sort seen in \fbqs\ is common,
then we predict that quasars of one subtype which are more reddened than average
for that subtype will be more likely to develop BAL troughs of the next subtype.
(Note that when its SDSS spectrum was taken, \fbqs\ itself had 
$\Delta(g-i)=0.33$, bluer than the average $\Delta(g-i)\simeq 0.5$
for the LoBAL quasars from Reichard et al. (2003).  $\Delta(g-i)$ is the $g-i$
colour of a quasar minus the average $g-i$ for quasars at the same redshift.)

\subsection{Average Episodic Lifetime of BAL Outflows} \label{tB}

We can use multiple-epoch spectroscopy of BAL (and non-BAL) quasars 
to constrain the average episodic lifetime of BAL outflows
by expanding upon \S 4.2 of Gibson et al. (2008).
We take all time intervals in the quasar rest frame.

We define an episodic lifetime as the time over which a quasar's spectrum
exhibits a BAL outflow, using a predetermined definition of what constitutes a
BAL trough.  
(Note that by this definition, BAL episodes do not overlap in time.)
The appearance and disappearance of a BAL in a spectrum thus define the start
and end of a BAL episode, regardless of whether they reflect
the creation and destruction of a BAL outflow or merely the passage of 
one or more structures in a BAL outflow across our line of sight.  
The episodic BAL lifetime is what we can observe, but it is worth remembering
that the lifetime of the entire BAL outflow must be equal to or longer than
the episodic lifetime while the lifetime of a structure in a BAL outflow
can be longer or shorter than the episodic lifetime (\S~\ref{tage}).

Suppose that quasars are sufficiently luminous to be included in a sample 
for an average time $t_Q$ (differential selection effects would admittedly
have to be considered unless bolometric luminosity selection could be used)
that a fraction $f_{LOS}$ of those quasars exhibit an average
of $N_B$ BAL episodes along our line of sight during their lifetimes
(meaning that a fraction $1-f_{LOS}$ never do so),
and that those BAL episodes have an average episodic lifetime $t_B$.
If we consider $t_B$ and $t_Q$ to be fixed,
although they are actually the means of distributions,
then the average instantaneous fraction of quasars showing BALs
along our line of sight is 
\begin{equation} \label{fBAL}
f_{BAL}=f_{LOS} N_B t_B / t_Q.
\end{equation}
Gibson et al. (2008) assumed $f_{LOS}=1$;
we relax that assumption to $f_{LOS}\gtrsim 0.2$, where the lower limit is set
by the observed BAL fraction $f_{obs}\simeq 0.2$ (e.g., Knigge et al. 2008).
Note that equation \ref{fBAL} does not depend on 
how BAL episodes are spaced over the quasar's lifetime,
as long as the episodic BAL lifetime $t_B$ 
does not change systematically over the quasar's lifetime.

Now consider two observations of a BAL
quasar separated by some (rest-frame) time interval.

The probability $p^-$ of observing a BAL disappear 
from a BAL quasar over a time interval $t^-_{obs} < t_B$ is 
\begin{equation} \label{p-}
p^- = t^-_{obs} / t_B
\end{equation}
if we also have $t^-_{obs} < t_{gap}$,
where $t_{gap}=(t_Q/N_B)-t_B$ is the average time between BAL episodes.

The probability $p^+$ of observing a BAL appear in a non-BAL quasar 
over a time interval $t^+_{obs}<t_B$ is conceptually similar.
It is the fraction of all non-BAL quasars which ever appear as a BAL quasar
($ { f_{LOS}-f_{BAL} \over 1-f_{BAL} } $)
times the number of BAL episodes that start during such quasars' observable
lifetimes ($\simeq N_B$) 
times the probability of a BAL episode starting during time
$t^+_{obs}$ ($\simeq t^+_{obs}/t_Q$).
To write a more exact expression for $p^+$, we must
account for the time non-BAL quasars which can appear as BAL quasars
spend as BAL quasars, and for the fraction $f_0$ of such quasars 
which have BALs at the start of their observable lifetimes.
The full derivation is given in the Appendix; 
the resulting expression for $p^+$ is:
\begin{equation} \label{p+}
p^+ = { f_{BAL} \over 1-f_{BAL} }{ t^+_{obs} \over t_B }
\left[ 1 - { f_0 \over N_B } \right].
\end{equation}
Neglecting the correction term in brackets 
(unity for $f_0=0$ or large $N_B$),
if all quasars appear as BAL quasars at some point ($f_{LOS}=1$)
then we have $f_{BAL}={N_Bt_B\over t_Q}$ (equation \ref{fBAL})
and we recover $ p^+ = { N_Bt^+_{obs}/t_Q \over 1-N_Bt_B/t_Q } $.  That
expression is the Gibson et al. (2008) approximation $p^+ = N_Bt^+_{obs}/t_Q$
with the quasar lifetime $t_Q$ replaced by the time a quasar spends as
a non-BAL quasar.

In practice, by constraining $p^-$ we constrain $t_B$,
and that constraint plus knowledge of $f_{BAL}$ constrains 
$f_{LOS}N_B/t_Q$ via equation \ref{fBAL}
and $f_0/N_B$ via equation \ref{p+}.
Limits on $t_Q$ can be invoked to constrain $f_{LOS}N_B$, but only in certain
cases might it be possible to constrain $f_0$, $N_B$ or $f_{LOS}$ separately
by monitoring the appearance or disappearance of BAL outflows (see below).
The assumptions underlying this analysis
are that fixed $t_Q$ and $t_B$ are reasonable approximations and
that no systematic change in $t_B$ occurs over the quasar's lifetime
(made to derive $f_{BAL}$), that 
the timescale $t^-_{obs}$ over which a BAL outflow is observed to disappear
obeys $t^-_{obs} < t_B$ and $t^-_{obs} < t_{gap}$ (made to derive $p^-$)
and that the timescale $t^-_{obs}$ over which a BAL outflow is observed to 
appear in a non-BAL quasar obeys $t^+_{obs} < t_B$ (made to derive $p^+$).
Future analyses of BAL episodic lifetimes could relax these assumptions
as well as consider BAL episodes that overlap in time, whether at the
same velocities or at different ones.

Meanwhile, it is instructive to calculate the constraints 
obtained by considering the BALs in FBQS J1408+3054 to have disappeared,
even though the \mgii\ BAL has not quite done so.  Those constraints
are presented in the following paragraphs, along with more conservative
constraints (in parentheses) obtained by considering FBQS J1408+3054 as a
case of extreme BAL variability, but not of BAL disappearance.

Of 13 LoBALs and FeLoBALs discovered in the FIRST Bright Quasar Survey
(Becker et al. 2000, 2001)
and with later spectra from the SDSS, only FBQS J1408+3054 shows dramatic
changes in the absorption covering fraction and equivalent width.
FBQS J1408+3054 is still the only such strongly variable BAL quasar
when the parent sample is expanded to include the 24 FBQS HiBAL quasars
with later spectra from the SDSS.
Using the binomial distribution,
this result translates to a constraint of 
$p^- = 0.027_{-0.024}^{+0.074}$
($p^- < 0.060$); 
all limits and ranges given in this section are 90\% confidence.
We have also inspected 156 FBQS non-LoBALs with later spectral coverage of
the \mgii\ region from the SDSS.  None of those quasars
exhibited \mgii\ BALs in their SDSS spectra,
yielding a constraint of $p^+ < 0.0147$.

Using that combined $p^-$, we constrain individual BAL lifetimes to be
$45 < t_B < 980$ years ($t_B > 60$ years), 
as compared to the Gibson et al. (2008) result of $t_B > 43$ years.
This range of lifetimes is broadly consistent with independently derived
age limits on the BAL structure in \fbqs\ (\S~\ref{Multi} and \S~\ref{tage}),
Taking $f_{BAL}=0.2$, the combined constraint on $p^+$ yields
the constraint $f_0/N_B > 0.5$ 
for $t_B=45$ years but no constraint ($f_0/N_B>0$) 
for any $t_B>97$ years.
No constraint on $f_0/N_B$ is possible if we consider FBQS J1408+3054 
to be a case of extreme BAL variability rather than of BAL disappearance.

Note that placing a sufficiently strong upper limit on the episodic
BAL lifetime $t_B$ has the potential to place interesting constraints
on $f_0/N_B$.  For example, suppose we have a constraint $t_B<86$ years 
from a $p^-=0.047_{-0.012}^{+0.015}$ measurement of 
14 BAL troughs in a sample of 300 disappearing 
over 3 years (a rate within the current 90\% confidence limits).
Then the current constraint of $p^+<0.0085$ would require $f_0/N_B>\frac{1}{9}$.
That would mean $N_B\leq 9$, as $f_0\leq 1$ by definition.
In turn, that would constrain $t_Q \leq 3800 f_{LOS}$ years,
or $t_Q \leq 3800$ years as $f_{LOS}\leq 1$ by definition as well.
(Such a lifetime is a factor of three smaller than the minimum episodic quasar
lifetime suggested by the proximity effect; see Martini 2004.)
In general, a limit on $t_Q$ can be set when $p^-$ is significantly larger
than $p^+$, as follows.  If BALs are seen to
disappear more frequently than they appear,
then maintaining a constant $f_{BAL}$ requires that a significant fraction of
quasars enter the sample as BAL quasars and that $N_B$ not be too large,
so that eliminating one BAL appearance per quasar has an appreciable impact.
Once upper limits are available for $N_B$ and $t_B$, an upper limit to $t_Q$
follows because the quasar lifetime cannot be arbitrarily larger
than $N_Bt_B$ while still maintaining the observed $f_{BAL}$.

\subsection{Implications for the FR~II-BAL Link}

\fbqs\ was identified as likely having an FR~II radio morphology 
by Gregg et al. (2006), who found that BAL quasars with FR~II morphologies
are roughly a factor of ten less common than would be expected if
FR~II morphologies and BAL absorption were independent phenomena.
\fbqs\ qualifies as a radio-loud FR~II quasar if the radio sources identified 
as its FR~II lobes are in fact associated with it, 
or as a radio-intermediate, core-dominated quasar if not (Gregg et al. 2006).

The weakening of the \mgii\ absorption in \fbqs\ slightly reduces the
significance of the anticorrelation between BAL strength (as measured by the
balnicity index) and radio-loudness parameter $R^*$ for FR~II BAL quasars
found by Gregg. et al. (2006).
However, it remains true that FR~II BAL quasars define the upper envelope
of that anticorrelation among all radio-selected BAL quasars.

The implications of the FR~II-BAL link for BAL quasar variability
depend on the reason(s) for the FR~II-BAL link.

There could be some (magneto)hydrodynamic mechanism which disfavors the 
contemporaneous production of strong BAL winds and 
strong radio jets that produce FR~II morphologies.
Such a mechanism could explain why HiBAL and LoBAL quasars are
underrepresented among quasars with $\log R^* > 2$ (White et al. 2007).
Any such mechanism might also have to explain why LoBAL quasars are
overrepresented among quasars with $1<\log R^*<2$ (White et al. 2007); 
however, no correction was made in that analysis for the 
substantially greater reddening of LoBAL quasars 
as compared to HiBAL quasars (Reichard et al. 2003).
In any case,
the transience and variability of BAL outflows as a function of $R^*$ in
this model would depend on the column density variations 
of the outflows as a function of $R^*$.

Alternatively,
what if the anticoincidence of BAL winds and FR~II radio morphologies occurs 
because the former tend to occur early in a quasar's lifetime and
the latter can occur only when a BAL shroud has been sufficiently reduced in
column density by the quasar's radiation pressure?
In such an `evolutionary' model,
quasars with larger lobe separations should on average be older and
nearer the ends of their BAL phases, resulting in greater transience and
variability of BAL outflows along a given sightline in such objects.

Observations of BAL variability and transience as a function of lobe
separation and $R^*$ are needed to test the evolutionary
model and provide constraints any dynamical model of BAL outflows must match.

\section{Conclusion} \label{con}

The quasar FBQS J1408+3054 was discovered as a spectacular FeLoBAL quasar,
and historical data indicate it had been one for $\gtrsim$20 rest-frame years
prior to its discovery, but it is now only a modestly absorbed LoBAL quasar.
Its \feii\ absorption outflowing at 12,000 km~s$^{-1}$ disappeared
over a rest-frame time span of between 0.6 and 5 years,
and its \mgii\ equivalent width
decreased by a factor of two over the same time period.
This variability was likely caused by a BAL structure moving out of our line of
sight to the ultraviolet continuum emitting region of the quasar's accretion
disk, indicating a transverse velocity between 2600 km~s$^{-1}$ and
22,000 km~s$^{-1}$.

In the context of a disk wind model, we connect the observed radial velocity
and variability timescale of the \feii\ BAL structure to its velocity vector
and distance from the quasar.  We have studied what constraints can be
placed on those quantities given the unknown inclination of the accretion
disk and the unknown angle between the wind velocity and the disk plane.
We estimate that the BAL structure which moved out of our line of sight in
\fbqs\ is located between 5800~$R_{Sch}$ and 46,000~$R_{Sch}$ from the black 
hole and that the velocity vector of the wind is oriented 
at $\sim$24$^\circ$ to our sightline.

There could be other BAL structures at different distances from \fbqs,
but the close correspondence of the outflow velocity of \mgii\ with 
that of \feii\ strongly suggests that the \mgii\ absorbing structure is
located at the same distance as that of \feii.  Continued monitoring
of \fbqs\ will be useful to constrain the size of the \mgii\ structure, should
it too move out of our line of sight, and to constrain the time required for
another \feii\ absorption structure to move into our line of sight.

Lastly, we have worked out in greater detail than heretofore
how multiple-epoch spectroscopy of BAL and non-BAL quasars 
can be used to constrain the average lifetime of BAL episodes, $t_B$,
and potentially the average number of BAL episodes per quasar.
At the moment, the 90\% confidence
limit on $t_B$ is $>$60 rest-frame years.
Future observations of BAL quasar trough variability should greatly
improve upon that constraint.

\section{Acknowledgments}

We thank the referee for a thoughtful review.
PBH and KA were supported by NSERC,
WNB by NASA ADP grant NNX10AC99G, DPS by NSF grant AST-060734 
and RRG by NASA Chandra grant AR9-0015X and NSF grant AST07-09394.  Some of
this work was performed under the auspices of the U.S. Department of Energy 
by Lawrence Livermore National Laboratory under Contract DE-AC52-07NA27344.
The Hobby-Eberly Telescope (HET) is a joint project of the University of Texas
at Austin, the Pennsylvania State University, Stanford University,
Ludwig-Maximillians-Universit\"at M\"unchen, and Georg-August-Universit\"at
G\"ottingen.  The HET is named in honor of its principal benefactors,
William P. Hobby and Robert E. Eberly.  The Marcario Low-Resolution
Spectrograph is named for Mike Marcario of High Lonesome Optics, who
fabricated several optics for the instrument but died before its completion;
it is a joint project of the Hobby-Eberly Telescope partnership and the
Instituto de Astronom\'{\i}a de la Universidad Nacional
Aut\'onoma de M\'exico.
Funding for the SDSS and SDSS-II (http://www.sdss.org) has been provided by the Alfred P. Sloan Foundation, the Participating Institutions, the National Science Foundation, the U.S. Department of Energy, the National Aeronautics and Space Administration, the Japanese Monbukagakusho, the Max Planck Society, and the Higher Education Funding Council for England.  The SDSS is managed by the Astrophysical Research Consortium for the Participating Institutions.

\appendix

\section{The Probability of Observing a BAL Outflow Appear}

Equation \ref{p+} is derived
as follows.  Consider a quasar of observable lifetime $t_Q$
which has $N_B$ BAL episodes during that lifetime, each of duration $t_B$
(thus the BAL episodes do not overlap).
We can observe the quasar as a non-BAL for a total time $ t_Q-N_Bt_B $.
Each BAL phase is preceded by a non-BAL phase (except in quasars which start
their observable lifetimes as BALs, for which we make a correction later).
Thus, $N_B t^+_{obs}$ is
the total amount of time which is $t^+_{obs}$ later than any non-BAL phase
and in which the quasar is a BAL.
The probability of observing a non-BAL as a BAL some time $t^+_{obs}$ later 
is the ratio of that amount of
time to the total amount of time which is $t^+_{obs}$ later than
any non-BAL phase: $ t_Q-N_Bt_B-\bar{x}t^+_{obs} \simeq t_Q-N_Bt_B $.
(The $x$ factor is zero for quasars which end their observable lifetimes as
BALs, and unity otherwise.  Its average value $\bar{x}$ depends on the relative
frequency of those occurrences, about which we do not wish to make any 
assumptions (though we note that for randomly occurring BAL episodes,
$\bar{x}=1-N_Bt_B/t_Q$).  Fortunately, our assumption that $t^+_{obs}<t_B$ 
makes it reasonable to neglect the term involving $\bar{x}$.)
Thus we write $p^+$ as:
\begin{equation}
p^+ = { f_{LOS}-f_{BAL} \over 1-f_{BAL} }{N_B t^+_{obs}/t_Q \over 1-N_Bt_B/t_Q}
\end{equation}
Replacing $f_{LOS}$ with $f_{BAL}t_Q \over N_Bt_B$ (equation \ref{fBAL})
and inserting a factor ${t_B\over t_B}$:
\begin{equation}
p^+ = { t^+_{obs} f_{BAL} \over 1-f_{BAL} } { N_B \over t_Q } {t_B \over t_B}
\left[ { t_Q \over N_B t_B } - 1 \right] \left[ {1 \over 1 - N_Bt_B/t_Q} \right]
\end{equation}
Multiplying $N_Bt_B \over t_Q$ through the first term in brackets:
\begin{equation}
p^+ = { t^+_{obs}f_{BAL}/t_B \over 1-f_{BAL} }
\left[ 1 - N_Bt_B/t_Q \right] \left[ {1 \over 1 - N_Bt_B/t_Q} \right]
\end{equation}
The terms in brackets cancel, leaving only the first term.  

Finally, consider a quasar which starts its observable lifetime as a BAL
quasar.  It will have only $N_B-1$ BAL episodes which could be observed
to start.  If a fraction $f_0$ of quasars which exhibit BALs at some point
start out as BAL quasars, then we have
\begin{equation}
p^+= {t^+_{obs}f_{BAL}/t_B \over 1-f_{BAL}} {f_0 (N_B-1) \over N_B}
+ {t^+_{obs}f_{BAL}/t_B \over 1-f_{BAL}} (1-f_0) {N_B \over N_B}
\end{equation}
\begin{equation}
p^+= {t^+_{obs}f_{BAL}/t_B \over 1-f_{BAL}} 
{ f_0N_B-f_0 + N_B-f_0N_B \over N_B }
\end{equation}
yielding equation \ref{p+} at last.

%

\end{document}